\documentclass[doublecol]{epl2}

\usepackage{epsfig}
\usepackage{amsmath,amssymb,wasysym,epsfig,capt-of,ifthen,calc}
\usepackage{bbm}
\usepackage{latexsym} 
\usepackage{setspace} 
\usepackage{array} 
\usepackage{delarray} 
\usepackage{afterpage} \usepackage{graphicx}
\usepackage{dcolumn}
\usepackage{bm}
\usepackage{float} \usepackage{supertabular} \usepackage{longtable}
\usepackage{amsmath}\bibstyle{apsrev}

\title{Percolation properties of growing networks under an Achlioptas process}
\shorttitle{Percolation properties of growing networks under an Achlioptas process}

\author{Su Do Yi\inst{1} \and Woo Seong Jo\inst{1} \and Beom Jun Kim\inst{1}
\and Seung-Woo Son\inst{2}}
\shortauthor{Su Do Yi \etal}

\institute{
  \inst{1} BK21 Physics Research Division and Department of Physics,
  Sungkyunkwan University, Suwon 440-746, Korea \\
  \inst{2} Department of Applied Physics, Hanyang University, Ansan 426-791, Korea
}


\abstract{ We study the percolation transition in growing networks under an Achlioptas process (AP). At each time step, a node is added in the network and, with the probability $\delta$, a link is formed between two nodes chosen by an AP. We find that there occurs the percolation transition with varying $\delta$ and the critical point $\delta_c=0.5149(1)$ is determined from the power-law behavior of order parameter and the crossing of the fourth-order cumulant at the critical point, also confirmed by the movement of the peak positions of the second largest cluster size to the $\delta_c$. Using the finite-size scaling analysis, we get $\beta/\bar{\nu}=0.20(1)$ and $1/\bar{\nu}=0.40(1)$, which implies $\beta \approx 1/2$ and $\bar{\nu} \approx 5/2$. The Fisher exponent $\tau = 2.24(1)$ for the cluster size distribution is obtained and shown to satisfy the hyperscaling relation. }

\begin{document}

\maketitle

\section{Introduction}

Percolation is one of the most frequently applied models to various natural phenomena in statistical physics~\cite{StaufferBook}. Nowadays, engaged in the complex network studies, percolation has been investigated in diverse network structures and has spun off many variations~\cite{Dorogovtsev2008}. In a complex network with growing number of links, soon after a certain critical number of links, a large cluster comparable to the system size $N$ does emerge. The percolation transition had been considered as {\em continuous} one not only in regular lattices, but also in complex networks. Erd\H{o}s and R\'{e}nny (ER) found that, in random networks, there is no giant cluster if the average degree $\langle k \rangle$ is less than $1$ in the limit of $N \rightarrow \infty$~\cite{BollobasBook}. However, the giant cluster emerges at $\langle k \rangle=1$ and its size monotonically increases as $\langle k \rangle$ increases with the order parameter critical exponent $\beta=1$.

Recently, it has been reported that the percolation transition becomes {\em explosive} when one incorporates a special rule, so-called Achlioptas process (AP), which suppresses the growth of large clusters while adding a link~\cite{Achlioptas2009}. The term `explosive' is used to emphasize the {\em discontinuous} emergence of the giant cluster as increasing the number of links. After the possibility of a discontinuous percolation transition had been reported in Ref.~\cite{Achlioptas2009}, there were several studies on explosive percolation in various systems such as a square lattice ~\cite{Ziff2009}, scale-free networks~\cite{Radicchi2009,Cho2009}, and with various rules~\cite{Cho2011}. Before long, however, later researches showed analytically and numerically that actually the percolation transition is still continuous, but very abrupt, which implies a very small value of $\beta$~\cite{daCosta2010,Grassberger2011,Lee2011,Riordan2011}.

In the real world, many networks evolve in time as varying not only the number of links, but also the number of nodes. Most of real networks are growing networks, contrary to static networks with fixed number of nodes~\cite{Barabasi1999,Dorogovtsev2000,Krapivsky2001}. In growing networks, the degree inhomogeneity caused by ages of nodes makes the percolation transition abnormal~\cite{Callaway2001,Dorogovstsev2001}. The percolation properties of growing networks show {\em infinite-order phase transition} like the Berezinskii-Kosterlitz-Thouless phase transition in condensed matter physics~\cite{Dorogovtsev2008}. The infinite-order phase transitions were observed in randomly growing networks~\cite{Callaway2001}, networks with preferential attachment process~\cite{Dorogovstsev2001}, growing trees~\cite{Hasegawa2010}, and protein interaction network models~\cite{JKim2002}. Here infinite-order transitions imply very smooth transition contrary to the abrupt transition of the explosive percolation. Here our question is what happens if we perform these two processes, growing and AP, together.

In this paper, we numerically investigate the percolation behavior on growing networks which is ruled by an AP. While growing a network under an AP, the smooth transition nature from the growing effect would compete with the nature to be abrupt transition from an AP. 

\section{Growing via an Achlioptas process}
\label{model}

Starting from a single node, at each time step, we add a new node to the network, and at the probability $\delta$, an undirected link is added. Instead of randomly adding a link to a pair of nodes, we choose a connection to delay the growth of large clusters, according to the AP~\cite{Achlioptas2009}. Here we follow the convention that double links and self-links are not allowed. Among various types of APs~\cite{Cho2011,daCosta2010,Grassberger2011,Lee2011}, we follow the da Costa model~\cite{daCosta2010} since it is expected to be more stringent selection of small clusters than other models still preserving the key feature of the phenomena of explosive percolation~\cite{Achlioptas2009}. To determine one end of the link, we choose two nodes uniformly at random and attach the end to the node that belongs to the smallest cluster. We repeat the same for the other end of the link.

In this growing model, the time $t$ obviously corresponds to the system size $N$, and the average degree $\langle k \rangle = 2L/N = 2 \delta$, where $L$ is the number of links. We observe the system in time and measurements are made when $N=32000$, $64000$, $...$, $1024000$, and $2048000$. Most of measures in this study are averaged over more than $1000$ ensembles, but we drop sample average notation $[ \ldots ]_{\text{sample}}$ only for convenience from now on. Note that the connection probability $\delta$ is the only parameter in our model, controlling the link density via $\delta=L/N$.

\begin{figure}[h]
\includegraphics[width=0.49\textwidth]{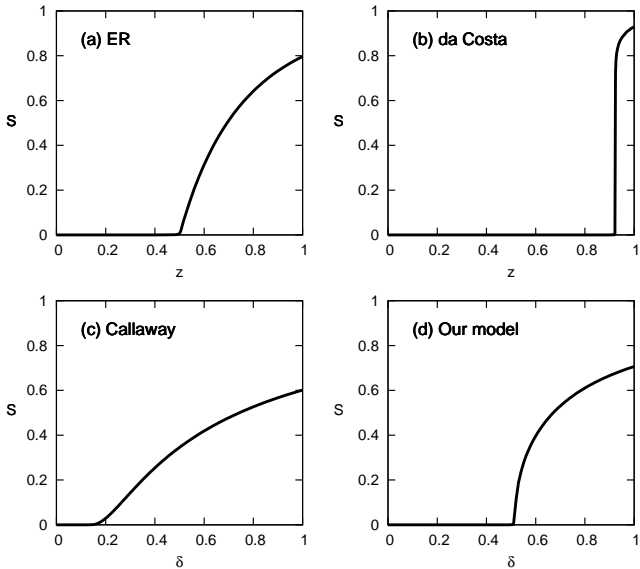}
\caption{
The fraction $S$ of the largest cluster as a function of the occupation density $z$ for (a) ER network and (b) network under AP in Ref.~\cite{daCosta2010}, and as a function of the connection probability $\delta$ for (c) randomly grown network in Ref.~\cite{Callaway2001} and (d) our model. For static and growing networks, the numbers of links correspond to $z$ and $\delta$, respectively. One can expect that the order parameter exponent $\beta$ for our model lies between (a) $\beta=1$ and (b) $\beta=0.0555(1)$. Also, it is clear that the nature of transition is not of infinite-order as shown in Callaway (c). All figures are for $N=2048000$ with $1000$ samples.
}
\label{fig:fig1}
\end{figure}

\section{Results}
\label{results}

First, we observe the fraction of the largest cluster, defined as the relative size of the giant cluster with respect to the system size $N$.  Figure~\ref{fig:fig1}(d) shows that our model exhibits a well-defined percolation transition as the the probability $\delta$ increases. Comparing with the transition point $\delta_c=1/8$ for the randomly grown model in~\cite{Callaway2001} [see Fig.~\ref{fig:fig1}(c)], we find that the AP suppresses the occurrence of the giant component and thus delays the transition toward larger $\delta_c$. However, it does not look like abrupt transition at all. It seems there is no explosive percolation transition in this model diluted by growing effect. In comparison with result from Ref.~\cite{daCosta2010} for a AP without growing [See Fig.~\ref{fig:fig1}(b)], the transition appears not as abrupt as explosive percolation. Furthermore, difference from Ref.~\cite{Callaway2001} [see Fig.~\ref{fig:fig1}(c)] appears to imply that our present model does not exhibit infinite-order transition. If we also compare with ER [Fig.~\ref{fig:fig1}(a)], in which $\beta=1$ is known, the seemingly diverging slope near $\delta_c$ implies that $\beta<1$. From the above comparisons with ER, da Costa, and Callaway (see Fig.~\ref{fig:fig1}), one can make conjecture that our model likely shows a continuous phase transition with $0<\beta<1$.

\begin{figure}[t]
\includegraphics[width=0.49\textwidth]{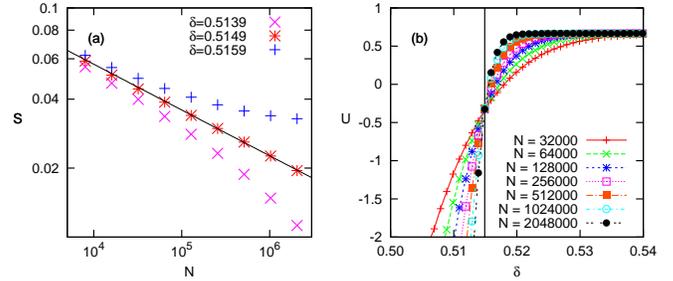}
\caption{(Color online) (a) Log-log plot the largest cluster sizes $S$ as a function of system size $N$ at different $\delta$ values. To find
$\delta_c$, we assume that $s$ depends on $N$ with the power-law of $S \sim N^{-\beta/\bar{\nu}}$.
We use the least-square fitting of log-log values and find that $\beta/\bar{\nu}=0.20(1)$ at $\delta=0.5149$.
(b) The fourth-order cumulant $U = 1-\langle S^4 \rangle / 3\langle S^2 \rangle ^2$ as a function of $\delta$.
The vertical line is for $\delta=0.5149$.
}
\label{fig:fig2}
\end{figure}

To find the critical point $\delta_c$ and the critical exponents, we assume the standard finite-size scaling (FSS) form for a continuous phase transition
\begin{equation}
S(\Delta,N)=N^{-\beta/\bar{\nu}} f(\Delta N^{1/\bar{\nu}}),
\end{equation}
where $\Delta = (\delta-\delta_c)/\delta_c$. If we use the correct $\delta_c$, $S$ must show the power-law behavior $N^{-\beta/\bar{\nu}}$. We investigate it for different
$\delta$ values [see Fig.~\ref{fig:fig2}(a)] and find that $\delta_c=0.5149(1)$.
From the least-square fitting, we find that the slope $\beta/\bar{\nu}=0.20(1)$.
We confirm the $\delta_c$ again with the fourth-order cumulant $U=1-\langle S^4 \rangle/ 3\langle S^2\rangle^2$. While there is no size dependency in FSS form, it should cross on $\delta_c$~\cite{Binder1981}.
As seen in Fig.~\ref{fig:fig2}(b), curves for different system sizes all cross at $\delta=0.5149$.

\begin{figure}[h]
\includegraphics[width=0.49\textwidth]{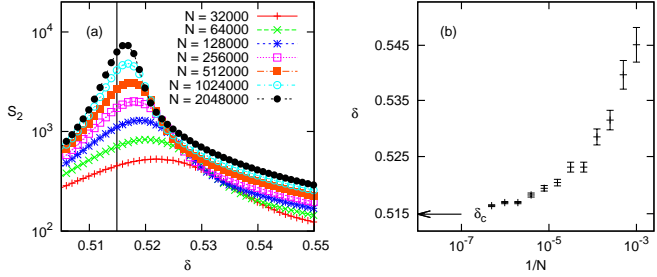}
\caption{(Color online)
(a) The second-largest cluster sizes $S_2$ as a function of $\delta$ for each $N$. The vertical line is for $\delta=0.5149$. (b) The peak positions in (a) as a function of $1/N$. The peak positions approaches $\delta_c$ from above as $N$ increases.
}
\label{fig:fig3}
\end{figure}

\begin{figure}[bh!]
\center
\includegraphics[width=0.4\textwidth]{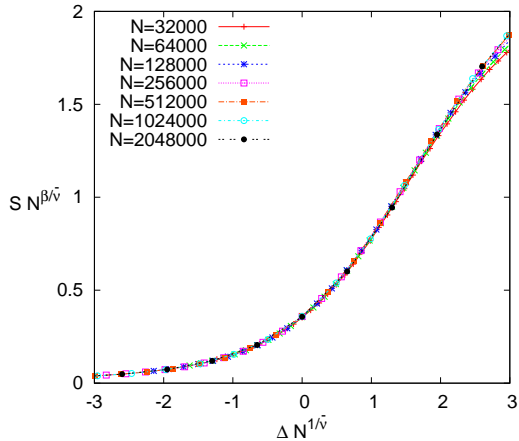}
\caption{(Color online) Finite-size scaling collapse using $S(\Delta,N)=
N^{-\beta/\bar{\nu}} f(\Delta N^{1/\bar{\nu}})$, where $\Delta = (\delta-\delta_c)/\delta_c$. It shows the best fit with $1/\bar{\nu}=0.40(1)$
when $\beta/\bar{\nu}=0.20(1)$, which implied $\beta \approx 1/2$ and $\bar{\nu} \approx 5/2$ in this model.
}
\label{fig:fig4}
\end{figure}
\begin{figure}[th!]
\center
\includegraphics[width=0.4\textwidth]{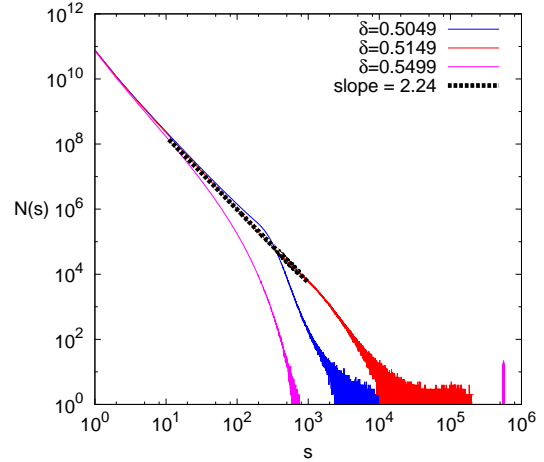}
\caption{(Color online) Number of clusters $N(s)$ of size $s$ for $100000$ realizations for subcritical, near-critical, and supercritical points. At the supposed critical point, the power-law distribution of clusters clearly appears in a broad range of $s$. The least-square fitting at $\delta_c$ gives us the slope of $\tau=2.24(1)$ as denoted by black dashed line.
}
\label{fig:fig5}
\end{figure}

Additionally, we compute the size of the second largest cluster $S_2$ to confirm the transition threshold. It is also a good indicator of the phase transition and have a tendency that the maximum peak position goes to the critical point as the system size becomes larger. It is clear that the peak positions moves toward $\delta_c$ as $N$ increases from Fig.~\ref{fig:fig3}(a) and (b). All these indicators clearly support $\delta_c=0.5149$.

Using $\beta/\bar{\nu}=0.20$ and $\delta_c=0.5149$, we try to make the curves for various system sizes collapse to one FSS form of $S$ as shown in Fig.~\ref{fig:fig4}. Since $\delta_c$ and $\beta/\bar{\nu}$ are already found in Fig.~\ref{fig:fig2}(a), the only tunable parameter is $\bar{\nu}$. From this analysis, we find $1/\bar{\nu}=0.40(1)$, which gives us the order parameter exponent $\beta \approx 1/2$, and the correlation volume exponent $\bar{\nu} \approx 5/2$. Interestingly, estimated values of $\beta \approx 1/2$ and $\bar{\nu} \approx 5/2$ are very close to the ones for the globally coupled Kuramoto model~\cite{Hong2007}. Whether the agreement is a mere coincidence or not, we have not been able to answer yet. Contrary to the percolation transition in randomly grown network~\cite{Callaway2001}, our present model does not show the infinite-order transition.

Furthermore, Fig.~\ref{fig:fig5} shows the number of clusters $N(s)$ as a function of cluster size $s$ at subcritical, near-critical, and supercritical regions. The distribution of the number of clusters $N(s)$ is quite different from each other with different $\delta$ as shown in Fig.~\ref{fig:fig5}. At subcriticality, there exists a bump and the rapid decay. Near criticality, $N(s)$ shows the power-law behavior as observed in the case of percolation in static network~\cite{Radicchi2009}. The Fisher exponent $\tau$ is given by
\begin{equation}
N(s) \sim s^{-\tau}
\end{equation}
for $s \gg1$. We get $\tau = 2.24(1)$ from the least-square fitting of log-log values in Fig.~\ref{fig:fig5}.
It is consistent with the result of the hyperscaling relation~\cite{Privman1990}
\begin{equation}
\frac{\beta}{\bar{\nu}} = \frac{\tau-2}{\tau-1},
\end{equation}
which yields $\tau = 2.25$. This confirms that the order parameter exponent and correlation volume exponent obtained from the FSS analysis are self-consistent. For $\delta > \delta_c$, $N(s)$ has the exponential cut-off at a certain size and there is a remote small peak which corresponds the giant cluster.

\section{Summary and Discussion}
\label{summary}

In summary, we have studied the phase transition in the growing network under an Achlioptas process (AP).
Similarly to static networks, AP makes the transition delayed toward larger $\delta_c$. Our main finding is that an AP changes the transition nature from the infinite-order transition to the second-order transition for a growing network. The critical point $\delta_c$ is estimated to be $0.5149(1)$ from the power-law decay of $S$ versus $N$ and from the finite-size scaling (FSS) analysis, which is confirmed from the crossing of the fourth-order cumulants and the movement of the peak positions of the second largest cluster size. Using the conventional FSS form for continuous transitions, we find that $\beta \approx 1/2$ and $\bar{\nu} \approx 5/2$. Additionally, the number of cluster $N(s)$ as a function of cluster size $s$ shows the power-law behavior at the critical point $\delta_c$. The Fisher exponent is in a good agreement with other exponents through the hyperscaling relation. We summarize our results in Table.~\ref{table}. One can see that, under random process, growing network has higher order of transition than static network. Under AP, $\beta$ for a growing network is larger than for a static network. That is, growth of network makes transition smoother than static case. Our results also suggest that an Aclioptas process makes the transition shaper than the random process.
\begin{table}[t]
\caption{Nature of a percolation transition. Regardless of link connection rules, growing networks exhibit smoother nature of transition (larger $\beta$). On the other hand, AP makes the transition sharper (smaller $\beta$) both for static and growing networks.}
\center
\begin{tabular}{|l|c|c|}
    \hline
     & static network & growing network \\
    \hline
    \small{Random} & \small{second-order} & \small{infinite-order}  \\
    \small{process} &  $\beta=1$ & $\beta \rightarrow \infty$ \\
    \hline
    \small{Achlioptas} & \small{abrupt transition} & \small{second-order} \\
    \small{process} & $\beta \ll 1 (\neq 0)$ & $\beta = 1/2$ \\
    \hline
\end{tabular}
\label{table}
\end{table}

\begin{acknowledgments}
We appreciate the initial idea of this problem and helpful comments from Maya Paczuski. This work was supported by Basic Science Research Program through the National Research Foundation of Korea (NRF) funded by the Ministry of Education, Science and Technology via No. 2012R1A1A1012150 (S.-W.S.) and No. 2010-0008758 (B.J.K.).
\end{acknowledgments}

\end{document}